\documentclass[prd,superscriptaddress,showpacs,nofootinbib,amsmath,amssymb,aps,11pt]{revtex4}

\usepackage{bm}
\usepackage{amsfonts}
\usepackage{latexsym}
\usepackage[latin1]{inputenc}
\usepackage{graphicx}
\usepackage{amsmath}
\usepackage{palatino}
\usepackage{mathpazo}
\usepackage{booktabs}
\usepackage{dcolumn}

\def\jnl@style{\it}
\def\aaref@jnl#1{{\jnl@style#1}}

\def\aaref@jnl#1{{\jnl@style#1}}

\def\aj{\aaref@jnl{AJ}}                   
\def\apj{\aaref@jnl{ApJ}}                 
\def\apjl{\aaref@jnl{ApJ}}                
\def\apjs{\aaref@jnl{ApJS}}               
\def\apss{\aaref@jnl{Ap\&SS}}             
\def\aap{\aaref@jnl{A\&A}}                
\def\aapr{\aaref@jnl{A\&A~Rev.}}          
\def\aaps{\aaref@jnl{A\&AS}}              
\def\mnras{\aaref@jnl{Mon.~Not.~Roy.~Astron.~Soc.}}             
\def\prd{\aaref@jnl{Phys.~Rev.~D}}        
\def\prc{\aaref@jnl{Phys.~Rev.~C}}  
\def\prl{\aaref@jnl{Phys.~Rev.~Lett.}}    
\def\qjras{\aaref@jnl{QJRAS}}             
\def\skytel{\aaref@jnl{S\&T}}             
\def\ssr{\aaref@jnl{Space~Sci.~Rev.}}     
\def\zap{\aaref@jnl{ZAp}}                 
\def\nat{\aaref@jnl{Nature}}              
\def\aplett{\aaref@jnl{Astrophys.~Lett.}} 
\def\apspr{\aaref@jnl{Astrophys.~Space~Phys.~Res.}} 
\def\physrep{\aaref@jnl{Phys.~Rep.}}      
\def\physscr{\aaref@jnl{Phys.~Scr}}       
\def\commat{\aaref@jnl{Comm.~Math.~Phys.}}              
\def\science{\aaref@jnl{Science}}               
\def\cqg{\aaref@jnl{Classical Quant.~Grav.}}            
\def\jpcs{\aaref@jnl{JPCS}}                                     
\def\ijmpd{\aaref@jnl{Int.~J.~Mod.~Phys.~D}}                    
\def\grg{\aaref@jnl{Gen.~Relat.~Gravit.}}               
\def\rpp{\aaref@jnl{Rep.~Prog.~Phys.}}          
\def\npa{\aaref@jnl{Nucl.~Phys.~A}}        
\def\lrr{\aaref@jnl{Living Rev.~Rel.}}                   
\def\jcap{\aaref@jnl{J.~Cosmology Astropart.~Phys.}}    
\def\rmp{\aaref@jnl{Rev.~Mod.~Phys.}}   

\allowdisplaybreaks[1]

\begin{document}

\title{A new type of  dark compact objects in massive tensor-multi-scalar theories of gravity}

\author{Stoytcho S. Yazadjiev}
\email{yazad@phys.uni-sofia.bg}
\affiliation{Theoretical Astrophysics, Eberhard Karls University of T\"ubingen, T\"ubingen 72076, Germany}
\affiliation{Department of Theoretical Physics, Faculty of Physics, Sofia University, Sofia 1164, Bulgaria}
\affiliation{Institute of Mathematics and Informatics, 	Bulgarian Academy of Sciences, 	Acad. G. Bonchev St. 8, Sofia 1113, Bulgaria}

\author{Daniela D. Doneva}
\email{daniela.doneva@uni-tuebingen.de}
\affiliation{Theoretical Astrophysics, Eberhard Karls University of T\"ubingen, T\"ubingen 72076, Germany}
\affiliation{INRNE - Bulgarian Academy of Sciences, 1784  Sofia, Bulgaria}


\begin{abstract}
In the present paper we consider special classes of tensor-multi-scalar theories of gravity whose target-space metric admits
Killing field(s) with a periodic flow. For such  tensor-multi-scalar theories we show that 
if the dynamics of the scalar fields is confined on the periodic orbits of the Killing field(s) then  
there exists a new type of compact objects -- the tensor-multi-scalar solitons formed by a condensation of the gravitational scalars. 
The existence of the tensor-multi-scalar solitons is proven by solving the fully non-linear eigenvalue problem which follows from the 
dimensional reduction of the field equations of  tensor-multi-scalar theories of gravity. 
The mass of the tensor-multi-scalar solitons can range at least from the mass of a neutron star to the mass of dark objects in the center of the galaxies in dependence of mass(es) of the gravitational scalars and which sector of massive scalars is excited. These facts show that the tensor-multi-scalar solitons could have  important implications for the dark matter problem. The existence of the tensor-multi-scalar solitons 
points towards the possibility that the dark matter, or part of it, is made of condensed gravitational scalars.   
    
\end{abstract}


\maketitle

\section{Introduction}

One of the most important and long standing problems in modern physics is the dark matter problem. Although a lot of efforts have been
devoted to this problem the nature of dark matter is still a mystery. There are many dark matter candidates, though currently there is no evidence for any of them. There are however strong indications that dark matter, or majority of it, is nonbaryonic in nature. 
Because dark matter has  been observed only indirectly via its gravitational field, it must barely interact with ordinary baryonic matter and electromagnetic radiation. This fact points towards a possible connection of  dark matter or part of it with fundamental fields 
which mediate  the gravitational interaction. Natural candidates of gravitational theories that could  directly be related to dark matter
are the tensor-multi-scalar theories \cite{Damour_1992}. In these theories the gravitational interaction is mediated not only by the spacetime metric but also by additional real scalar fields. In the physical Jordan frame these gravitational scalar fields do not interact directly with the baryonic matter and with the electromagnetic field, i.e. in certain sense they are dark in nature. Hence it is natural to assume that
there is a possible connection between the gravitational scalars and  dark matter (or part of it).        

In the present paper we show that in certain classes of tensor-multi-scalar theories if we allow the gravitational scalars to be dynamical and if their dynamics in the target-space manifold is properly constrained, then there exist 
a new type of compact objects which are soliton-like solutions to the  vacuum field equations of the tensor-multi-scalar theories. We call these solutions {\it tensor-multi-scalar solitons}. Since there is no direct interaction between the gravitational scalars and the electromagnetic field the tensor-multi-scalar solitons are dark in nature. In agreement with the present day observations their mass can range at least from the mass of a neutron star  to the mass of dark objects in the center of the galaxies in dependence of mass(es) of the gravitational scalars and
which massive sector is exited. These facts show that the tensor-multi-scalar solitons could have  important implications for the dark matter problem.  The existence of the tensor-multi-scalar solitons points towards the possibility that the  dark matter or part of it is made up by condensed gravitational scalars.

\section{Tensor-multi-scalar-theories and conserved currents}
In the tensor-multi-scalar theories of gravity \cite{Damour_1992,Horbatsch_2015}, as we mentioned, the gravitational interaction is mediated by the spacetime metric $g_{\mu\nu}$ and $N$ real scalar fields $\varphi^{a}$  which take value in a coordinate patch of an N-dimenional Riemannian target-space manifold ${\cal E}_{N}$ supplemented with (positively definite) metric $\gamma_{ab}(\varphi)$. In the Einstein frame the general action of the  tensor-multi-scalar theories of gravity is given by 
\begin{eqnarray}\label{Action}
S= \frac{1}{16\pi G_{*}}\int d^4\sqrt{-g}\left[R - 2g^{\mu\nu}\gamma_{ab}(\varphi)\nabla_{\mu}\varphi^{a}\nabla_{\nu}\varphi^{b} - 4V(\varphi)\right]  + S_{matter}(A^{2}(\varphi) g_{\mu\nu}, \Psi_{matter})
\end{eqnarray}
where $G_{*}$ is the bare gravitational constant, $\nabla_{\mu}$ and $R$ are the covariant derivative  and the Ricci scalar curvature with respect to  the Einstein frame metric $g_{\mu\nu}$, and $V(\varphi)\ge 0$ is the potential of the scalar fields. In order for the weak equivalence principle to be satisfied the matter fields, denoted collectively by $\Psi_{matter}$, are coupled only to the physical Jordan metric ${\tilde g}_{\mu\nu}= A^2(\varphi) g_{\mu\nu}$ where  $A^2(\varphi)$ is the conformal factor relating the Einstein and the Jordan metrics, and which, together with $\gamma_{ab}(\varphi)$ and $V(\varphi)$, specifies the tensor-multi-scalar theory.  

The Einstein frame field equations corresponding to the action (\ref{Action}) are the following
\begin{eqnarray}\label{FE}
&&R_{\mu\nu}= 2\gamma_{ab}(\varphi) \nabla_{\mu}\varphi^a\nabla_{\nu}\varphi^b + 2V(\varphi)g_{\mu\nu} + 8\pi G_{*} \left(T_{\mu\nu} - \frac{1}{2}T g_{\mu\nu}\right), \\
&&\nabla_{\mu}\nabla^{\mu}\varphi^a = - \gamma^{a}_{\, bc}(\varphi)g^{\mu\nu}\nabla_{\mu}\varphi^b\nabla_{\nu}\varphi^c 
+ \gamma^{ab}(\varphi) \frac{\partial V(\varphi)}{\partial\varphi^{b}} - 
4\pi G_{*}\gamma^{ab}(\varphi)\frac{\partial\ln A(\varphi)}{\partial\varphi^{b}}T, \nonumber
\end{eqnarray}
where $T_{\mu\nu}$ is the Einstein frame energy-momentum tensor of matter and $T$ is its trace. $\gamma^{a}_{\, bc}(\varphi)$ are 
the Christoffel symbols with respect to the target-space metric $\gamma_{ab}(\varphi)$. We shall consider here only the vacuum case with $T_{\mu\nu}=0$.

In order to avoid unnecessary technical complications we shall focus on the $N=2$ case. Nevertheless our approach presented below is general and, up to some technical details, is applicable to the case with arbitrary $N$. We shall consider a special class of tensor-multi-scalar theories which is defined as follows. We require that the metric $\gamma_{ab}(\varphi)$ admits a Killing field  $K^{a}$ with  a periodic flow and also, $A(\varphi)$ and $V(\varphi)$  be invariant under the flow of the Killing field $K^{a}$, i.e. ${\cal L}_{K} A(\varphi)=K^a\partial_{a}A(\varphi)=0$  and ${\cal L}_{K}V(\varphi)= K^a\partial_{a}V(\varphi)=0$. The existence of a such Killing field gives rise to a conserved current\footnote{It is woth noting that the conserved current $J^\mu$ exists even in the presence of matter when our requirements are satisfied.} $J^{\mu}$ which in the Einstein frame is given by 
\begin{eqnarray}\label{EFC} 
J^{\mu}= \frac{1}{4\pi G_{*}} g^{\mu\nu} K_{a}\partial_{\nu}\varphi^{a}.
\end{eqnarray}
Its Jordan frame version is
\begin{eqnarray}\label{JFC}
{\tilde J}^{\mu}= \frac{1}{4\pi G(\varphi)} {\tilde g}^{\mu\nu} K_{a}\partial_{\nu}\varphi^{a}
\end{eqnarray}
where $G(\varphi)= G_{*}A^2(\varphi)$. The fact that $J^{\mu}$ and ${\tilde J}^{\mu}$  are conserved currents, i.e. $\nabla_{\mu}J^{\mu}={\tilde \nabla}_{\mu}{\tilde J}^{\mu}=0$, follows from the fact that $K^a$  is a Killing field for $\gamma_{ab}(\varphi)$, the  field equations (\ref{FE}) and from ${\cal L}_K A(\varphi)={\cal L}_{K}V(\varphi)=0$.

A more explicit description  of our requirements is the following. Since the target-space manifold is 2-dimensional, there exists a coordinate patch, the so-called isothermal coordinates, in which the metric $\gamma_{ab}(\varphi)$ can  be written in the form 
\begin{eqnarray}
\gamma_{ab}(\varphi)= \Omega^2(\varphi)\delta_{ab}.
\end{eqnarray} 

In these coordinates the Killig field with the periodic orbits is explicitly given by $K=\varphi^{2}\frac{\partial}{\partial\varphi^1} - \varphi^{1}\frac{\partial}{\partial\varphi^2}$. In order for $K$ to be a Killing field  $\Omega$ must depend on $\varphi^a$ through the combination $\psi$ where $\psi^2=\delta_{ab}\varphi^a\varphi^b$, i.e. $\Omega=\Omega(\psi)$. The same applies to $A(\varphi)$ and $V(\varphi)$, i.e.
$A=A(\psi)$ and $V=V(\psi)$. 

In the present paper, in order to be more specific, we shall focus on maximally symmetric ${\cal E}_2$. 
In this case the isothermal coordinates are globally defined and
 \begin{eqnarray}
\Omega^2= \frac{1}{\left(1+ \frac{\kappa}{4}\delta_{ab}\varphi^a\varphi^b\right)^2} =\frac{1}{\left(1+ \frac{\kappa}{4}\psi^2\right)^2}
\end{eqnarray} 
where the constant $\kappa$ is the Gauss curvature of ${\cal E}_2$. The spherical geometry  corresponds to $\kappa>0$, the hyperbolic geometry 
to $\kappa<0$ while the flat geometry is given by $\kappa=0$.

\section{Tensor-multi-scalar solitons}

\subsection{Basic equations}

 The main question we want to answer is whether there exist strictly static (in both Einstein and Jordan frame), completely regular and asymptotically flat solutions to the vacuum equations of the tensor-multi-scalar theories with localized energy, i.e. soliton type solutions. By strictly static spacetimes we mean spacetimes  admitting everywhere timelike hypersurface orthogonal Killing field $\xi$. If the gravitational scalars $\varphi^a$ inherit the static symmetry of spacetime, i.e. the Lie derivative ${\cal L}_{\xi}\varphi^a=0$, then no nontrivial soliton solutions can exist \cite{Heusler_95}. In the present paper we show that in certain classes of tensor-multi-scalar theories if we allow the gravitational scalars to be dynamical and if their dynamics in the target-space manifold is properly constrained, then there exist soliton solutions to the vacuum field equations of the tensor-multi-scalar theories. As mentioned above we call these solutions {\it tensor-multi-scalar solitons}.

 A necessary condition for the existence of solitons is ${\cal L}_{\xi}\varphi^{a}\ne 0$ or in simple words the scalar fields (or part of them in the general case) should be dynamical. From a physical point of view in order for the solitons to exists the scalar field dynamics should be periodic in time and this is naturally  satisfied if the dynamics of the scalar fields is ``confined'' on the periodic orbits of the Killing field $K^a$. In geometrical terms this can be expressed in the following way 
\begin{eqnarray}\label{periodic}
{\cal L}_{\xi}\varphi^a= \omega K^a 	
\end{eqnarray}	
where $\omega$ is nonzero (real) constant.  With this condition imposed one can check that effective energy-momentum tensor $T^{(\varphi)}_{\mu\nu}=(4\pi G_{*})^{-1}\left[\gamma_{ab}(\varphi) (\nabla_{\mu}\varphi^a\nabla_{\nu}\varphi^b -\frac{1}{2}
g_{\mu\nu}\nabla_{\sigma}\varphi^a\nabla^{\sigma}\varphi^b) - V(\varphi)g_{\mu\nu}\right]$ of the gravitational scalars is static, i.e. ${\cal L}_{\xi}T^{(\varphi)}_{\mu\nu}=0$. One more consistency condition that must be satisfied  is the Ricci staticity condition $R[\xi]\wedge \hat{\xi}=0$ where $R[\xi]=\xi^{\mu}R_{\mu\nu}dx^\nu$ is the Ricci one-form and $\hat{\xi}=\xi_\mu dx^\mu$ is the Killing one-form naturally corresponding to the Killing field $\xi$. In view of (\ref{periodic}) the Ricci staticity condition reduces to 
\begin{eqnarray}\label{staticity}
{\hat J}\wedge {\hat \xi} = 0
\end{eqnarray}
with ${\hat J}=J_{\mu}dx^\mu$. We have to mention that our requirements automatically ensure that
the Jordan frame metric ${\tilde g}_{\mu\nu}=A^2(\varphi) g_{\mu\nu}$ is also static. This follows from the fact that  conformal factor $A(\varphi)$ is  static. Indeed we have ${\cal L}_{\xi} A(\varphi)= \partial_a A(\varphi){\cal L}_{\xi}\varphi^a=\omega K^a \partial_{a}A(\varphi)=0$.

In general not all solutions of (\ref{periodic}) satisfy (\ref{staticity}).
A physically natural solution to (\ref{periodic}), for which the condition (\ref{staticity}) is also satisfied, is given by 
\begin{eqnarray}
(\varphi^1, \varphi^2) = (\psi(x^i) cos(\omega t), \psi(x^i) \sin(\omega t)),
\end{eqnarray}
where $t$ is the time coordinate adapted to the Killing field $\xi$, i.e. $\xi=\frac{\partial}{\partial t}$, and $x^i$ are coordinates defined on the hypersurface $\Sigma$ orthogonal to $\xi$.

In what follows we will focus on static, spherically symmetric and asymptotically flat spactimes. In this case the spacetime metric takes 
the usual form 
\begin{eqnarray}
ds^2= - e^{2\Phi(r)}dt^2 + e^{2\Lambda(r)}dr^2 + r^2(d\theta^2 + \sin^2\theta d\phi^2)
\end{eqnarray} 
where $t$ is the time coordinate adapted to the Killing field $\xi$ and the scalar field $\psi$ depends on the radial coordinate $r$ only,
$\psi=\psi(r)$.  Taking into account all the above constructions, the dimensionally reduced vacuum field equations are the following 
\begin{eqnarray}\label{DRE}
&&\frac{2}{r}e^{-2\Lambda} \Lambda^\prime + \frac{1}{r^2}\left(1- e^{-2\Lambda}\right)= \Omega^{2}(\psi)\left[\omega^2e^{-2\Phi}\psi^2 + 
(\psi^\prime)^2e^{-2\Lambda}\right] + 2V(\psi), \nonumber \\
&&\frac{2}{r}e^{-2\Lambda} \Phi^\prime - \frac{1}{r^2}\left(1- e^{-2\Lambda}\right)= \Omega^{2}(\psi)\left[\omega^2e^{-2\Phi}\psi^2 + 
(\psi^\prime)^2e^{-2\Lambda}\right] - 2V(\psi),\\
&&\psi^{\prime\prime} + \left(\Phi^\prime - \Lambda^\prime  + \frac{2}{r}\right)\psi^\prime  \notag \\ 
&&\;\;\;\;\; +\left[ \omega^2e^{-2\Phi}\left(1 + 2\frac{\partial\ln\Omega}{\partial\psi^2}\psi^2\right) + 2\frac{\partial\ln\Omega}{\partial\psi^2}(\psi^\prime)^2 e^{-2\Lambda} - 2\Omega^{-2}(\psi) \frac{\partial V(\psi)}{\partial\psi^2}\right]e^{2\Lambda}\psi=0.
\nonumber
\end{eqnarray}

It is worth commenting on an alternative description/interpretation of our construction. Under the assumptions we made 
the Einstein frame field equations of the $N=2$ tensor-multi-scalar theories are formally equivalent to the Einstein equations coupled to a (massive and self-interacting) complex scalar field  $\Psi=\varphi^1 + i\varphi^2=\psi e^{i\omega t}$ with an exotic non-minimal kinetic term. From this formal point of view the tensor-multi-scalar solitons in the Einstein frame can be viewed as some sort of boson stars made of an exotic complex field. In this description/interpretation the conserved current  (\ref{EFC})  could be interpreted as a Noether current associated with the global $U(1)$ gauge transformations $\Psi \to \Psi e^{i\alpha}$ with $\alpha$ being a constant. The description/interpretation in terms of a complex field is appealing from a physical point, however it also suffers from serious drawbacks. This description is specific for $N=2$ and there is no natural extensions to higher dimensions with $N\ge 3$ in the general case. When we consider the presence of matter the description/interpretation in terms of one (or more) complex scalar field(s) only is either  impossible or rather problematic  in the general case  with $N\ge 3$ \cite{Doneva_2019}. That is why we prefer to use here the general geometrical description/construction instead of  introducing an effective exotic complex field.            

The conserved current (\ref{EFC})  leads to the conserved charge $q$ given by 
\begin{eqnarray}
q= -\int_{\Sigma} J^{\mu} \frac{\xi_{\mu}}{\sqrt{-g(\xi,\xi)}} \sqrt{h}d^3x,
\end{eqnarray}   
where $h$ is the determinant of the metric $h_{\mu\nu}$  induced on $\Sigma$. It should be noted that the conserved charges in the Einstein and in the Jordan frame are the same, $q={\tilde q}$. In the spherically symmetric case under consideration we have 
\begin{eqnarray}
q=\frac{1}{ G_{*}} \int^{+\infty}_{0} (\omega e^{-\Phi})\Omega^2(\psi)\psi^2 e^{\Lambda}r^2dr.
\end{eqnarray}

\subsection{Numerical results}

Let us return to the system (\ref{DRE}) of coupled nonlinear ordinary  differential equations. The natural boundary conditions for
(\ref{DRE}) are as follows. The asymptotic flatness requires $\Phi(\infty)=\Lambda(\infty)=\psi(\infty)=0$. The absence of a conical singularity at the center $r=0$ imposes $\Lambda(0)=0$. Regularity at the center requires $\psi^\prime(0)=0$. We also specify $\psi(0)=\psi_c$. With these boundary conditions the system (\ref{DRE}) forms a non-linear eigenvalue problem for $\omega$. Even more, in the present paper we will be interested only in zeronodes solutions for $\psi(r)$, i.e. in the lowest eigenvalue $\omega$.  We solved numerically the eigenvalue problem
for the following potential consistent with our requirements 
\begin{eqnarray}
V(\varphi)= V(\psi)= 
\frac{1}{2}m^2\psi^2 + \frac{1}{4}\lambda_{(4)} \psi^4 ,
\end{eqnarray}
where $m$ is the mass of the scalar fields and $\lambda_{(4)}$ is parameter governing the $\psi^4$-self-interaction of the scalar fields. 
This Einstein frame potential is natural from physical point of view and in addition it is universal for all tensor-multi-scalar theories, i.e. it is independent from $A(\varphi)$.  It is also more convenient to consider  the conserved charge $Q=m q$ instead of $q$. 
In presenting the numerical results it is convenient to use the normalized mass $M/(\frac{M^2_{Pl}}{2m})$ and the normalized 
conserved charge $Q/(\frac{M^2_{Pl}}{m})$ where $M_{Pl}$ is the Planck mass. The mass $M$ of the tensor-multi-scalar solitons is defined as the ADM mass in the Einstein frame. Since the scalar fields are massive and drop off  exponentially at infinity the
ADM masses in the Jordan and the Einstein frame coincide.

\begin{figure}
	\includegraphics[width=0.45\textwidth]{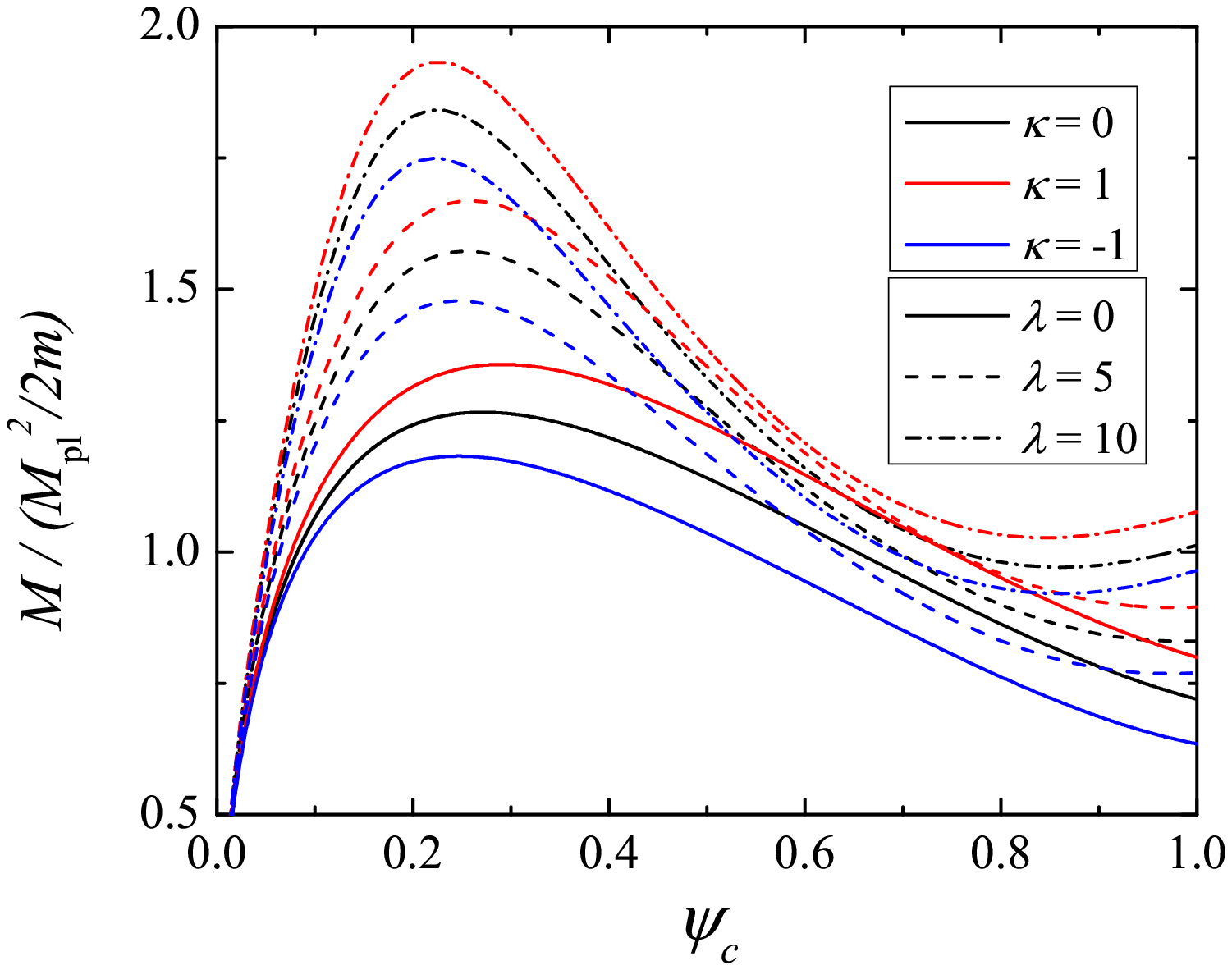}
	\includegraphics[width=0.45\textwidth]{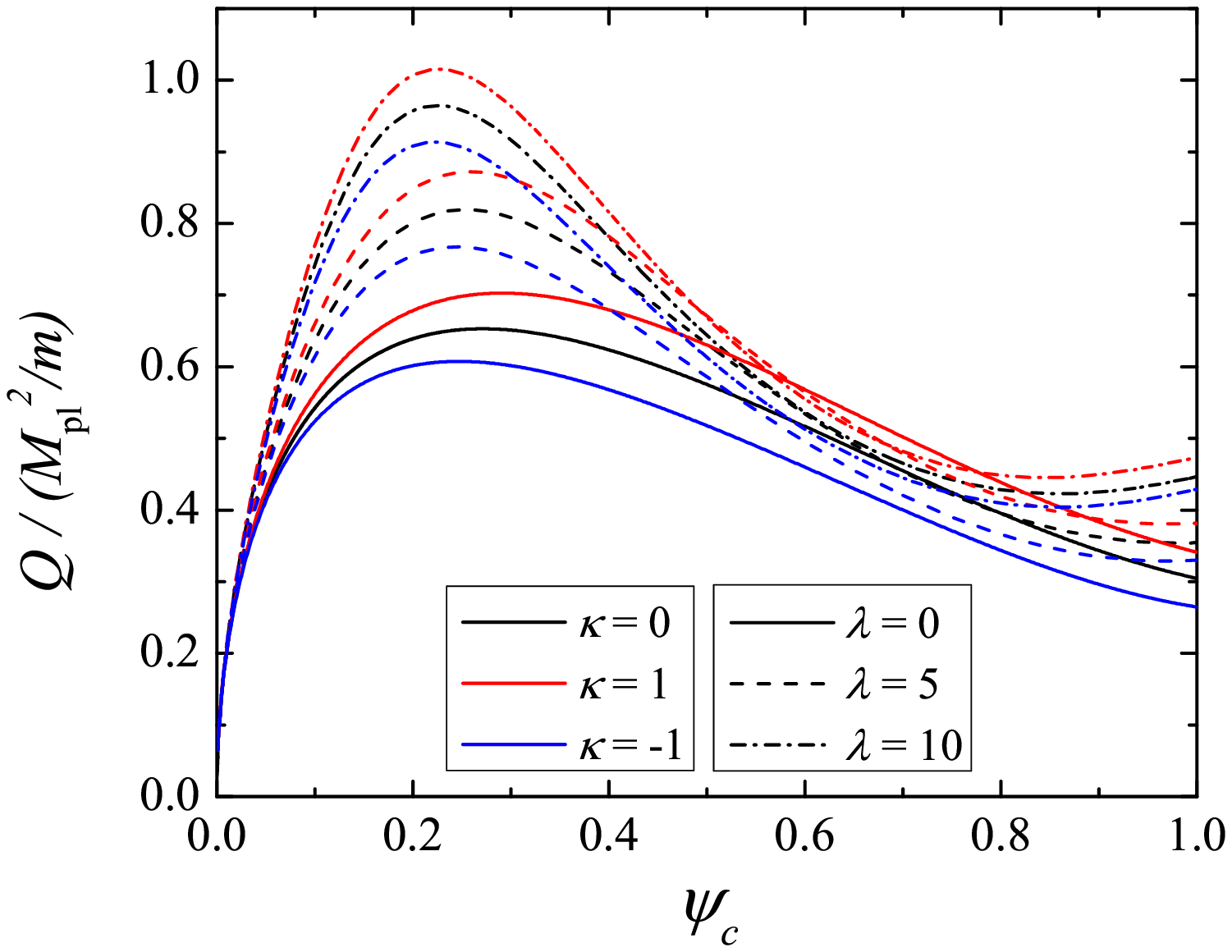}
	\caption{The  normalized mass $M$ (left panel) and the normalized conserved charge $Q$ (right panel) as functions of the central value of the scalar field $\psi_c$. Results for three values of $\lambda=\lambda_{(4)}/m^2$ ($\lambda=0,\;5,\;10$) and for $\kappa=0,\; \pm 1$ are presented in the graphs.	}
	\label{fig:MQ_psic}
\end{figure}

\begin{figure}
	\includegraphics[width=0.45\textwidth]{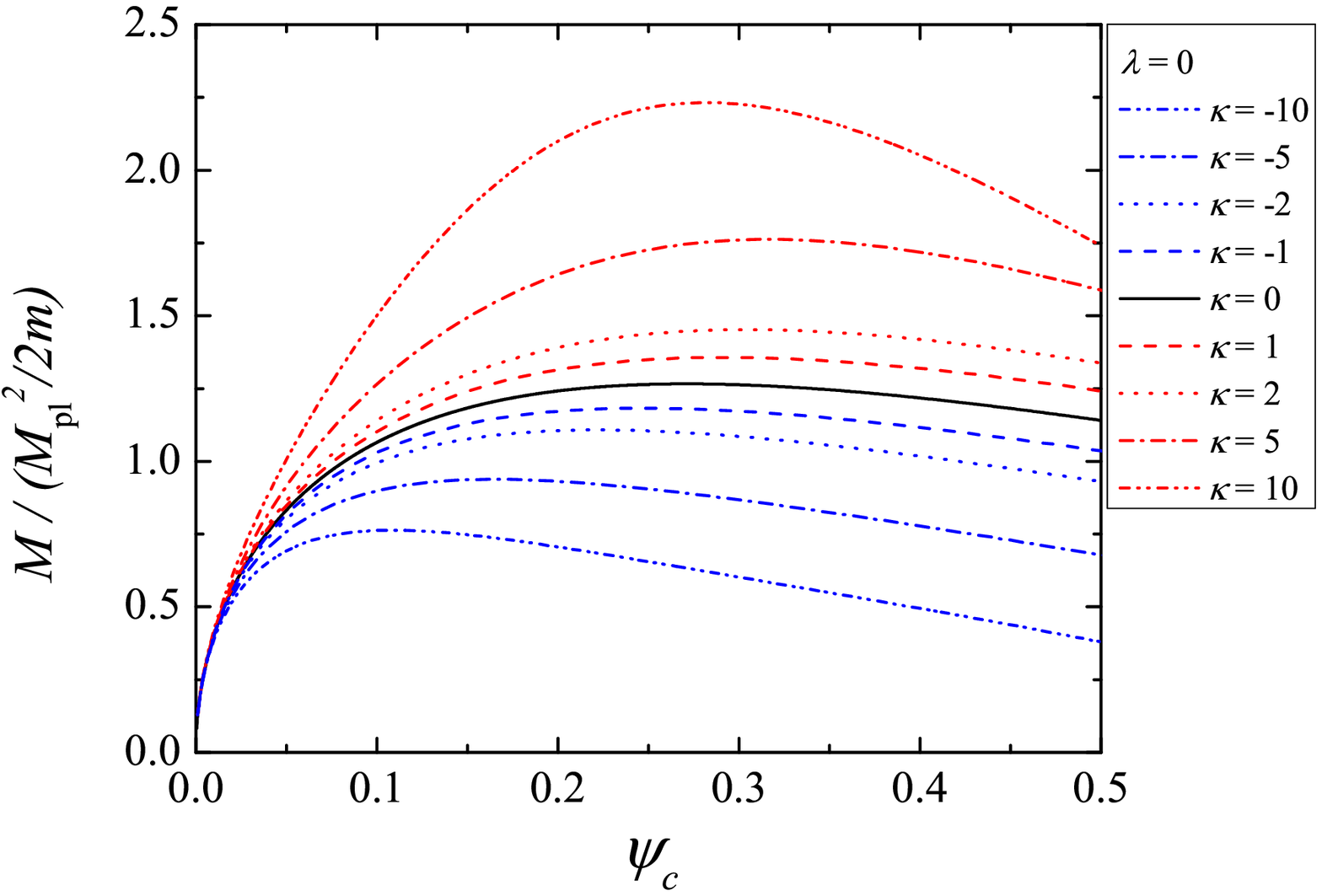}
	\includegraphics[width=0.45\textwidth]{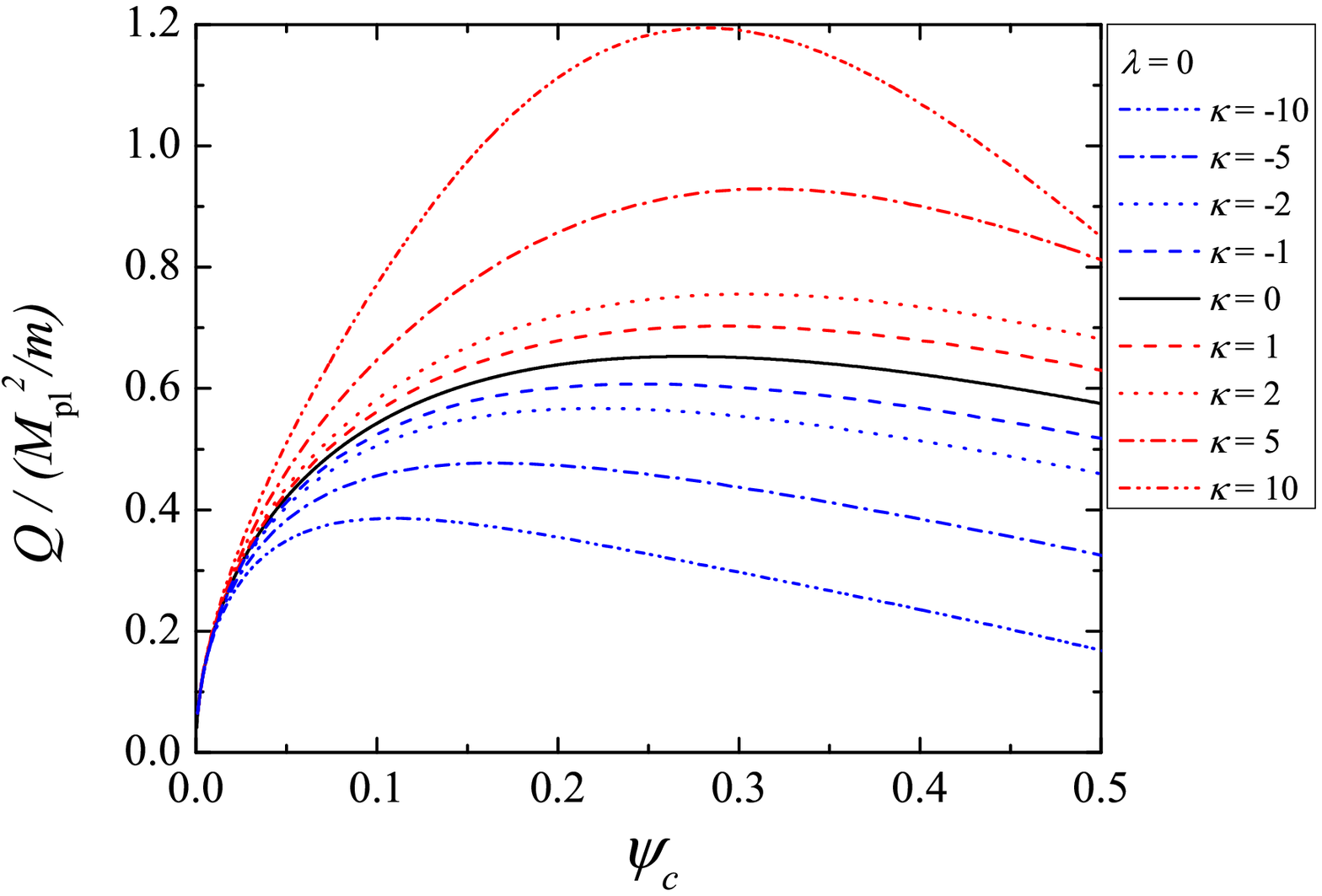}
	\caption{The normalized mass $M$ (left panel) and the normalized conserved charge $Q$ (right panel) as functions of the central value of the scalar field $\psi_c$ for several values of $\kappa$ are plotted. The parameter $\lambda$ is fixed to $\lambda=0$. 	}
	\label{fig:MQ_psic_vary_kappa}
\end{figure}

In Fig. \ref{fig:MQ_psic}  we present the normalized mass and the normalized  conserved charge of the tensor-multi-scalar solitons as    
a function of the central value of the scalar field $\psi_c$ for three values of the parameter $\lambda=\lambda_{(4)}/m^2$ and for three values of the curvature 
$\kappa$ representing the flat geometry ($\kappa=0$), spherical geometry ($\kappa=1$) and the hyperbolic geometry ($\kappa=-1$). 
From these results we can conclude that tensor-multi-scalar theories with positively curved target space produce solitons with 
the largest masses. The increase of the target space curvature  $\kappa$ leads to an increase of the soliton mass $M$ and the conserved charge $Q$ while the maximum mass decreases for  the hyperbolic geometry with $\kappa=-1$. The increase of the self-interaction parameter $\lambda$, on the other hands, leads to an increase of the mass and the charge for fixed $\psi_c$. In Fig. \ref{fig:MQ_psic_vary_kappa} we have studied in more detail the effect of varying $\kappa$ for fixed $\lambda=0$. The qualitative behavior is similar to the case of $\kappa=\pm 1$ in Fig. \ref{fig:MQ_psic} -- for positive values of $\kappa$ the mass and the charge increases while negative $\kappa$ lead to a decrease of $M$ and $Q$. One can also notice that for negative $\kappa$ the maximum of the mass happens at decreasing $\psi_c$ as $\kappa$ decreases. Assuming that the stable solutions are the ones for smaller $\psi_c$ before the first maximum of the mass is reached, this means that for $\kappa<0$ the range of central values of the scalar field where stable solutions exists shrinks with the decrease of $\kappa$.

Up to now all our considerations and results were general and independent from the specific tensor-multi-scalar theory. 
In order to calculate the physical (Jordan frame) radius of the solutions we have to specify the tensor-multi-scalar theory.
Our choice of a tensor-multi-scalar theory here is conservative and is just for illustration, namely 
\begin{eqnarray}\label{SST}
A(\varphi)= \exp(\frac{1}{2}\beta\psi^2)
\end{eqnarray}    
with $\beta=-6$.   The Jordan frame radius of the solitons is defined by  
\begin{eqnarray}
{\tilde R}_S= \frac{1}{ qG_{*}} \int^{+\infty}_{0} r A(\varphi) (\omega e^{-\Phi})\Omega^2(\psi)\psi^2 e^{\Lambda}r^2dr.
\end{eqnarray}   

\begin{figure}
	\includegraphics[width=0.45\textwidth]{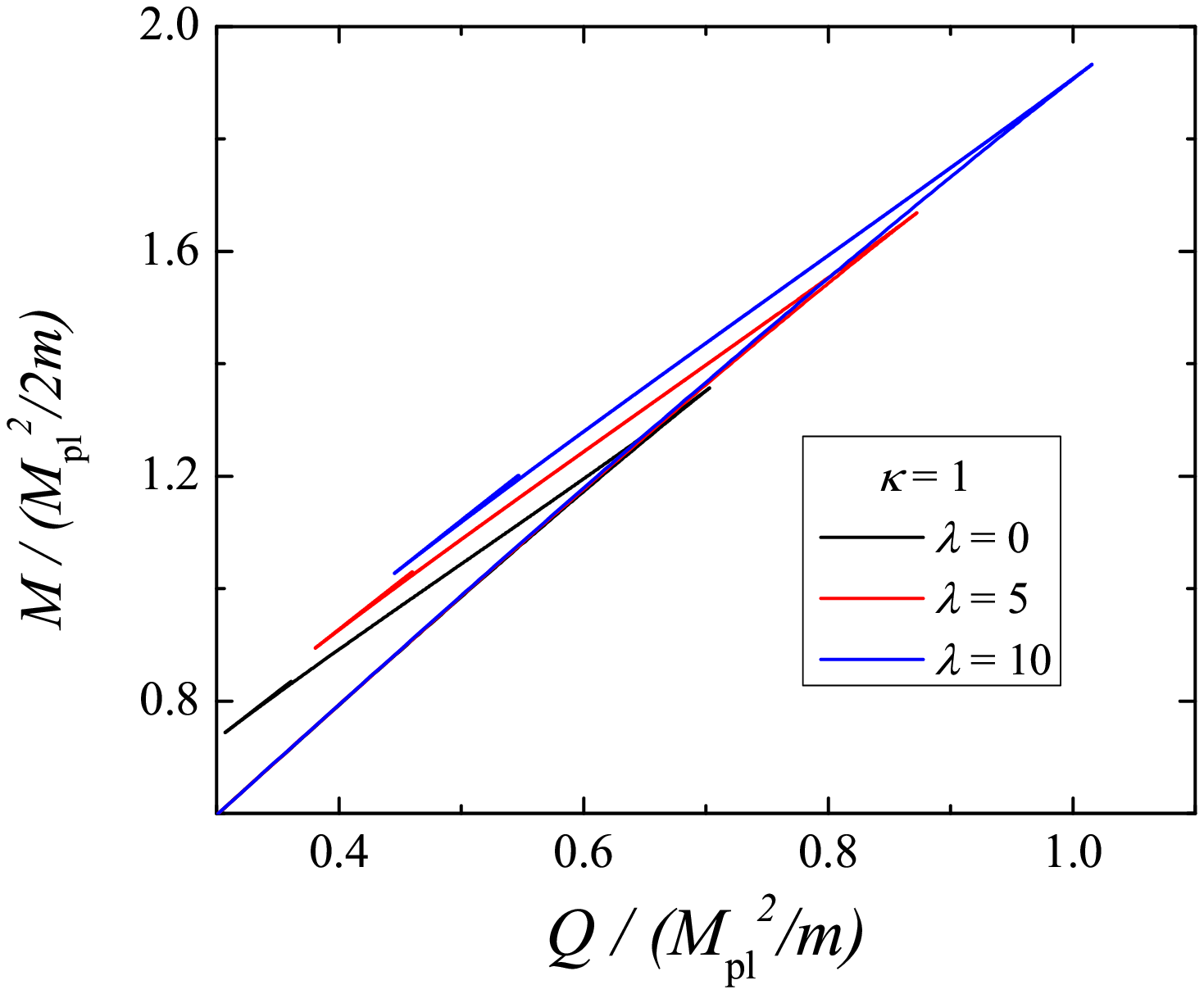}
	\includegraphics[width=0.45\textwidth]{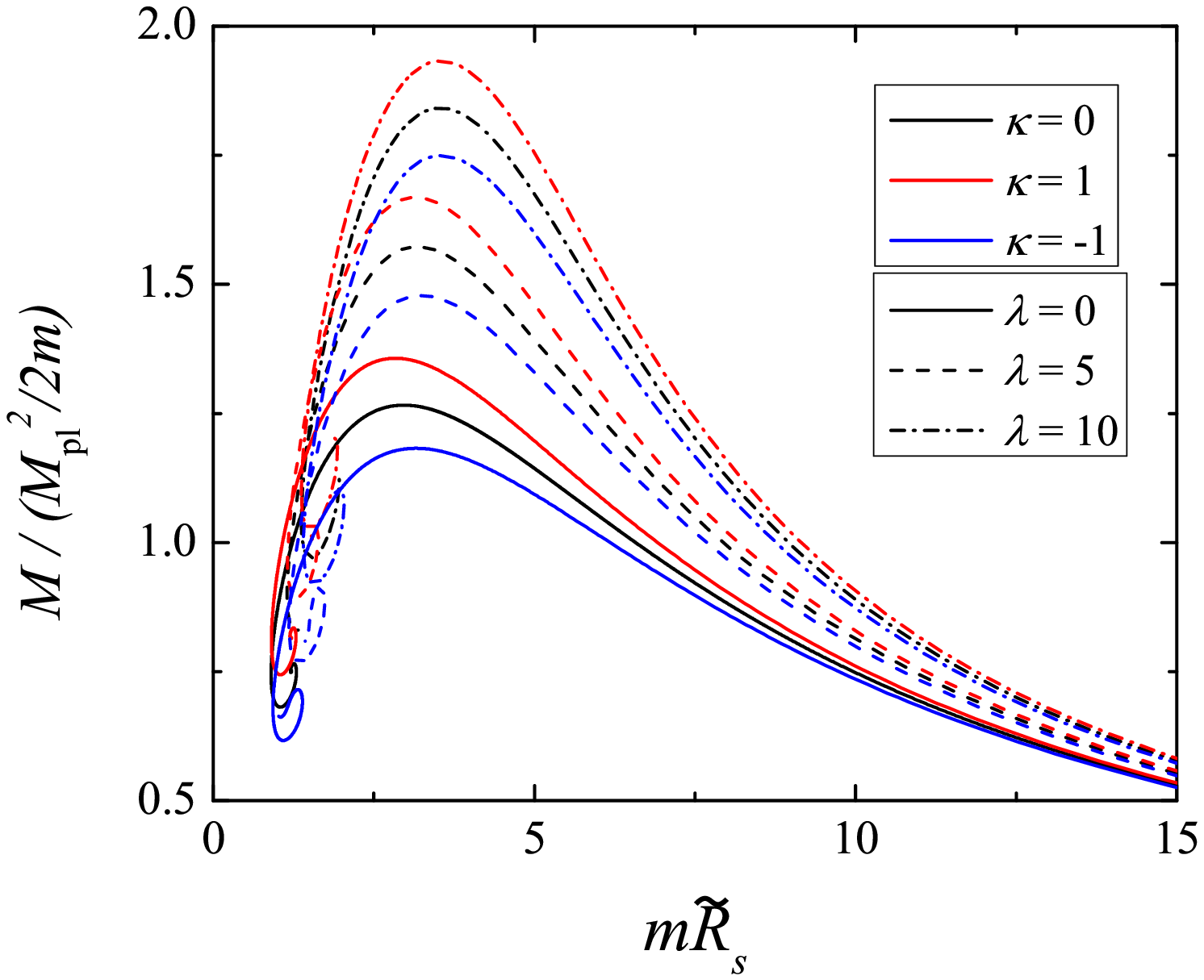}
	\caption{(left panel) The normalized mass as a function of the normalized conserved charge for $\kappa=1$ and several values of lambda. In this figure only several representative lines are presented compared to the other figures, in order to have a better resolution of the cusps at the turning points of the mass.(right panel) The  normalized mass $M$ as a function of the normalized radius $m {\tilde R}_s$. The notations are the same as in Fig. \ref{fig:MQ_psic}. 	}
	\label{fig:MQRs}
\end{figure}

In the right panel of Fig. \ref{fig:MQRs} the mass as a function of the normalized Jordan frame radius $m{\tilde R}_S$ is plotted. Results for three values of $\lambda$ and for $\kappa=0,\pm1$ are shown. As expected, the qualitative behavior is similar to the one observed in Fig. \ref{fig:MQ_psic} -- the increase of $\lambda$ and $\kappa$ leads to an increase of the mass. The maximum mass is also shifted to larger radii with the increase of $\lambda$. 

We have observed more than one extrema of the mass for increasing $\psi_c$ and every extremum is supposed to be connected with a change of stability of the solutions. Indeed, as one can see in the left panel of Fig. \ref{fig:MQRs} where the mass as a function of the conserved charge is plotted, the extrema of the mass correspond to cusps in the $M(Q)$ diagram. In this figure only one representative value of $\kappa=1$ is plotted in order to have better visibility of the cusps, but the qualitative behavior is the same for all other calculated values of $\kappa$. This leads to the conclusion that most probably the branch of solutions before the first maximum of the mass is reached, is stable. Of course a more rigorous answer can be given only after the linear stability of the solutions is examined and such a study is underway.

So far the mass of the scalar fields $m$ was arbitrary. In order to get some  insight into the values of the mass $M$ of the tensor-multi-scalar solitons  we have to constrain the possible values of $m$ from the observations. The possible constraints on $m$ depend  very strongly on the particular scalar-tensor theory, i.e. on $(\gamma_{ab}(\varphi), V(\varphi),A(\varphi))$.
Binary pulsars provide some of the tightest current constraints on scalar-tensor theories of gravity. However, the tensor-multi-scalar solitons
exist for all tensor-multi-scalar theories even for those where the scalar fields are not excited neither in neutron stars nor in the weak field limit. For such theories no  constraints can be imposed by the observations. A simple example of such a theory is the massive  theory $A(\varphi)=e^{\beta\psi^n}$ with $n\ge 3$ and even the massive theory (\ref{SST}) with $\beta>0$ when the equation of state of the neutron star matter satisfies the natural condition $\rho>3p$. In these cases the mass $M$ of the tensor-multi-scalar solitons can range in an extremely 
wide interval reaching even the typical masses of the galaxies if we assume very small values of $m$ of the order of $10^{-23}eV$.        

For some particular theories and under additional requirements a rough estimate of the soliton mass $M$ is possible. Such a theory is (\ref{SST}) with $\beta<0$  and  $\lambda_{(4)}=\kappa=0$. In this particular case, provided that the neutron stars are scalarized, a rough constraint on the mass $m$  can be obtained as in \cite{Ramazanoglu_2016, Yazadjiev_2016}, namely $10^{-16}eV<m<10^{-9}eV$. The lower bound guaranties that emitted scalar radiation is negligible while the upper bound  guaranties the mass term does not prevent the scalarization of the neutron star. Then for a theory under consideration we find $0.1M_{\odot}<M< 10^6 M_{\odot}$. When the self-interaction is switched on in our theory, the deviation from GR decreases if one increases the value for the parameter $\lambda_{(4)}$ in the self-interaction term. This can reconcile even wider range of values for the scalar-field mass $m$ with the observations \cite{Staykov_2018}.     

It is worth mentioning that when $N$ is large enough and $\gamma_{ab}(\varphi)$ admits several Killing fields $K_{L}$ with periodic orbits 
then the different $\psi_L$ can be characterized by different masses $m_{L}$. This in turn means that the tensor-multi-scalar solitons
can have various masses $M$ in dependence of which massive $L$-sector is excited.  

The above considerations show that, in agreement with the present day observations, in the general case the mass of the tensor-multi-scalar solitons  can range at least from the mass of a neutron star  to the mass of dark objects in the center of the galaxies, and even to mass of  the galaxies in dependence of mass(es) of the gravitational scalars and which massive sector is excited.

\section{Conclusion}

In the present paper we proved numerically the existence of a new type of compact objects within the framework of 
certain classes of tensor-multi-scalar theories of gravity  whose target space metric admits
Killing field(s) with a periodic flow.  These new compact objects, called  tensor-milti-scalar solitons, are soliton solutions made of condensed gravitational scalars. Since there is no direct interaction between the gravitational scalars and the electromagnetic field
the tensor-multi-scalar solitons are dark in nature. In agreement with the present day observations their mass can range at least from the mass of a neutron star  to the mass of dark objects in the center of the galaxies (and even more) in dependence of mass(es) of the gravitational scalars and which massive sector is excited. These facts show that the tensor-multi-scalar solitons could have  important implications for the dark matter problem.  The existence of the tensor-multi-scalar solitons  points towards the possibility that the dark matter is made of  condensed gravitational scalars. This possibility is very intriguing and will be investigated further in future publications. Since the ordinary matter is also a source of the gravitational scalars, in some cases one could expect a  nontrivial correlations between the ordinary matter and the dark matter described by the tensor-multi-scalar solitons.     

Finishing let us mention that our investigations show that there also exist  mixed configurations of tensor-multi-scalar solitons and 
relativistic (neutron) stars \cite{Doneva_2019}. Even more, our preliminary studies show that there exist mixed configurations of rotating (Kerr-like) black holes and tensor-multi-scalar solitons  which are characterized by their mass, their angular momentum and the conserved charge associated with the conserved current (\ref{EFC}) \cite{Doneva_2019a}, similar to the black holes reported in \cite{Herdeiro_2014}.

\section*{Acknowledgements}
DD would like to thank the European Social Fund, the Ministry of Science, Research and the Arts Baden-Wurttemberg for the support. DD is indebted to the Baden-Wurttemberg Stiftung for the financial support of this research project by the Eliteprogramme for Postdocs.
SY acknowledges financial support by the Bulgarian NSF Grant KP-06-H28/7. Networking support by the COST Actions  CA16104 and CA16214 is also gratefully acknowledged.



\end{document}